\documentclass[prb,aps,eqsecnum,showpacs,twocolumn]{revtex4}

\usepackage[dvips]{graphics}
\usepackage{epsfig}
\usepackage{graphicx}
\usepackage{amssymb}
\usepackage{bm}
\DeclareGraphicsExtensions{.bmp,.eps,.gif,.png}

\begin{document}

\title{Josephson coupling through ferromagnetic heterojunctions with non-collinear magnetizations}
\author{Z. Pajovi\'c}
\affiliation{Department of Physics, University of Belgrade, P.O.
Box 368, 11001 Belgrade, Serbia}

\author{M. Bo\v{z}ovi\'{c}}
\affiliation{Department of Physics, University of Belgrade, P.O.
Box 368, 11001 Belgrade, Serbia}

\author{Z. Radovi\'c}
\affiliation{Department of Physics, University of Belgrade, P.O.
Box 368, 11001 Belgrade, Serbia}

\author{J. Cayssol}
\affiliation{Universit$\acute{e}$ Bordeaux I, CPMOH, UMR 5798,
33405 Talence, France}

\author{A. Buzdin}
\affiliation{Universit$\acute{e}$ Bordeaux I, CPMOH, UMR 5798,
33405 Talence, France}

\begin{abstract}
We study the Josephson effect in clean heterojunctions that
consist of superconductors connected through two metallic
ferromagnets with insulating interfaces. We solve the scattering
problem based on the Bogoliubov--de Gennes equation for any
relative orientation of in-plane magnetizations, arbitrary
transparency of interfaces, and mismatch of Fermi wave vectors.
Both spin singlet and triplet superconducting correlations are
taken into account, and the Josephson current is calculated as a
function of the ferromagnetic layers thicknesses and of the angle
$\alpha$ between their magnetizations. We find that the critical
Josephson current $I_c$ is a monotonic function of $\alpha$ when
the junction is far enough from $0-\pi$ transitions. This holds
when ferromagnets are relatively weak. For stronger ferromagnets,
variation of $\alpha$ induces switching between $0$ and $\pi$
states and $I_c(\alpha)$ is non-monotonic function, displaying
characteristic dips at the transitions. However, the
non-monotonicity is the effect of a weaker influence of the
exchange potential in the case of non-parallel magnetizations. No
substantial impact of spin-triplet superconducting correlations on
the Josephson current has been found in the clean limit.
Experimental control of the critical current and $0-\pi$
transitions by varying the angle between magnetizations is
suggested.
\end{abstract}

\pacs{74.50.+r, 74.45.+c}

\maketitle

\section{Introduction}

The interplay between ferromagnetism and superconductivity in
superconductor (S)~--~ferromagnet (F) hybrid structures attracts
considerable interest for some time
already.\cite{reviewsSFS,reviewsSNS,Pokrovski,reviewbergeret}
Variety of interesting theoretical predictions, such as the
existence of $\pi$-state superconductivity in SF
multilayers,\cite{Bulaevski,FFLO,buzdin82,Radovic91,buzdin91,Demler,Tagirov_C,FCG,Bagrets}
has been confirmed
experimentally.\cite{Jiang,kontos01,ryazanov01,Obi,Lazar,Garifullin,zdravkov}
In SFS Josephson junctions, the Zeeman effect induces both a
strong decay and oscillations of superconducting correlations
inside the ferromagnet.\cite{ryazanov04,Valls01} The corresponding
oscillations of the Josephson critical current with thickness of
the ferromagnetic layer or the exchange energy have been
calculated in both the clean and the dirty limit in the framework
of quasiclassical theory of
superconductivity.\cite{buzdin82,Radovic91,buzdin91} Experimental
evidence for such oscillating behavior in weak ferromagnetic
alloys came only recently.\cite{ryazanov01,kontos01,zdravkov}
Similar experiments with strongly spin-polarized
ferromagnets\cite{blum02,blamire05} have been performed in
spintronic setup.\cite{zutic04} However, for strong ferromagnets
and finite transparency of interfaces the quasiclassical treatment
is no longer valid and Bogoliubov--de Gennes or Gor'kov equations
have to be solved.\cite{radovic03,cayssol04}

Heterostructures with superconductors coupled through
inhomogeneous ferromagnets have been also extensively
studied.\cite{bergeret01prltriplet,bergeret01AP,bergeret03prltriplet,bergeret03prbtriplet,kadigrobov,Eschrig03,blanter4,Champel}
The simplest example of such a structure contains a two-domain
ferromagnet with collinear magnetizations, either parallel (P) or
antiparallel (AP). In this case, besides spin singlet
superconducting correlations, only triplet correlations with zero
total spin projection exist.\cite{golubov2,You,lof} These
correlations penetrate into the ferromagnet over a short length
scale determined by the exchange energy. For non-collinear
magnetizations, triplet correlations with nonzero spin projection
are present as well; they are not suppressed by the exchange
energy, and consequently they are
long-ranged.\cite{reviewbergeret} It is predicted that triplet
components should have a dramatic impact on the Josephson effect,
displayed through a non-monotonic dependence of the critical
current on angle between
magnetizations.\cite{bergeret03prltriplet,bergeret01AP,bergeret03prbtriplet}
The triplet correlations were proposed as a possible explanation
for recent observations of a Josephson current through
half-metallic barriers.\cite{Keizer,Penya} In diffusive Josephson
junctions, the length scales associated with short- and long-range
correlations are, respectively, $\xi_{f}=\sqrt{\hbar D_f/h_{0}}$
and $\xi_{s}=\sqrt{\hbar D_f/k_{\rm B}T}$, where $D_f$ is the
diffusion constant in the ferromagnet, and thermal energy $k_{\rm
B}T$ is typically much smaller than the exchange energy $h_0$. It
is desirable to know if such interesting effects are also relevant
in ballistic SF heterostructures, where the ferromagnet coherence
length $\xi _{f}=\hbar v_{\rm F}^{(f)}/h_{0}$ is the only
characteristic length, $v_{\rm F}^{(f)}$ being the Fermi velocity.

In this paper we investigate the Josephson effect in a clean
SI$_1$F$_1$I$_2$F$_2$I$_3$S heterojunction where the ferromagnetic
interlayer consists of two mono-domain layers having a relative
angle $\alpha$ between their in-plane magnetizations. Using the
Bogoliubov--de Gennes formalism, we calculate the Josephson
current and demonstrate that its critical value $I_{c}$ is a
monotonic function of angle $\alpha$ when the junction is far
enough from $0-\pi$ transitions. However, this is possible when
ferromagnets are weak. For stronger ferromagnets, $I_{c}(\alpha)$
is a non-monotonic function of $\alpha$ with characteristic dips
related to the onset of $0-\pi$ transitions. This non-monotonicity
is a simple consequence of a weaker influence of the exchange
potential in the case of non-parallel magnetizations. No
substantial impact of spin-triplet superconducting correlations on
the Josephson current has been found in the clean limit. In fully
transparent symmetric junctions the ferromagnetic influence is
practically cancelled out for antiparallel magnetizations. This is
not the case for finite transparency of interfaces when coherent
geometrical oscillations of the critical Josephson current are
superimposed on oscillations related to the transitions between
$0$ and $\pi$ states. In the limiting case of zero exchange energy
our solutions reduce to the results for
SI$_1$N$_1$I$_2$N$_2$I$_3$S junctions with complex
non-ferromagnetic normal-metal (N) interlayer, previously studied
in the quasiclassical approach.\cite{galaktionov}

The paper is organized as follows. In Sec.~II we present the model
and solutions of the scattering problem  used for calculation of
the Josephson current. In Sec.~III we discuss numerical results
for the current-phase relation, and dependence of critical current
on angle between magnetizations and thickness of ferromagnetic
layers. We illustrate these results for various spin polarizations
and transparencies of the interfaces. Concluding remarks are given
in Sec.~IV.

\section{Model and solution}

\begin{figure}[h]
\begin{center}
    \includegraphics[width=7cm]{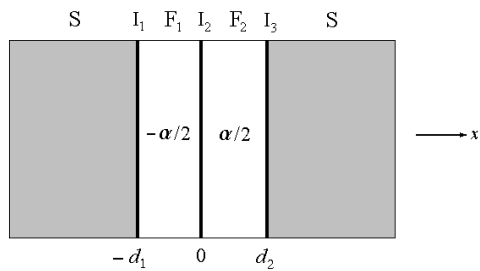}
\caption{Schematics of an SI$_1$F$_1$I$_2$F$_2$I$_3$S
heterojunction. The magnetization vectors lie in the $y$-$z$ plane
and form the opposite angles $\pm \alpha/2$ with respect to the
$z$-axis.} \label{Fig1}
\end{center}
\end{figure}

We consider a clean SI$_1$F$_1$I$_2$F$_2$I$_3$S heterojunction
consisting of superconductors (S), two uniform mono-domain
ferromagnetic layers (F$_{1}$ and F$_{2}$) with misorientation
angle $\alpha$ between their magnetizations, and nonmagnetic
interfacial potential barriers between metallic layers
(I$_1$--I$_3$), see Fig.~\ref{Fig1}. Superconductors are described
in the framework of standard BCS formalism, while for ferromagnets
we use the Stoner model with an exchange energy shift
$2h{(}\mathbf{r)}$ between the spin subbands.

Electron-like and hole-like quasiparticles with energy $E$
and spin projection $\sigma =\uparrow ,\downarrow $ are described by wavefunctions $%
u_{\sigma }(\mathbf{r})$ and $v_{\sigma }(\mathbf{r})$, where
$\mathbf{r}$ is the spatial coordinate. Using the four-component wave function $\Psi (\mathbf{r}%
)=[u_{\uparrow }(\mathbf{r}),u_{\downarrow }(\mathbf{r}),v_{\downarrow }(%
\mathbf{r}),v_{\uparrow }(\mathbf{r})]^{\rm T}$, we write the
Bogoliubov--de Gennes equation as
\begin{equation}
\check{\mathcal{H}}\Psi (\mathbf{r})=~E\Psi (\mathbf{r}),
\label{bgd0}
\end{equation}%
where
\begin{equation}
\check{\mathcal{H}}=\left(
\begin{array}{cc}
\widehat{H}_{+}(\mathbf{r}) & \widehat{\Delta }(\mathbf{r}) \\
\widehat{\Delta }^{\ast }(\mathbf{r}) & -\widehat{H}_{-}(\mathbf{r})%
\end{array}%
\right),   \label{bdg1}
\end{equation}%
while the $2\times 2$ blocks are given by $\widehat{H}_{\pm }(\mathbf{r})=H_{0}(\mathbf{r})\widehat{\bf 1}-h{(}%
\mathbf{r)}\sin {\alpha (}\mathbf{r)}\widehat{{\sigma }}_{2}\mp h{(}\mathbf{%
r)}\cos {\alpha (}\mathbf{r)}\widehat{{\sigma }}_{3}$ and $\widehat{\Delta }(%
\mathbf{r})=\Delta (\mathbf{r})\widehat{\bf 1}$. Here,
$\widehat{{\sigma}}_{i}$ and $\widehat{\bf 1}$ are the Pauli and
unity matrix, respectively, and $H_{0}(\mathbf{r})=-\hbar
^{2}\nabla ^{2}/2m+W(\mathbf{r})+U(\mathbf{r})-\mu $. The chemical
potential is denoted by $\mu $, $W(\mathbf{r})=\sum_i W_{i}\delta
(x-x_{i})$ is potential of the barriers at interfaces, and
$U(\mathbf{r})$ is the electrostatic potential. The $x$-axis is
chosen to be perpendicular to the layers, whereas $x_{1}=-d_{1}$,
$x_{2}=0$, and $x_{3}=d_{2}$ are coordinates of the interfaces.
The electron effective mass $m$ is assumed to be the same
throughout the layers. The difference $\mu-U(\mathbf{r})$ is equal
to the Fermi energy of superconductors, $E_{\rm F}^{(s)}$, or the
mean Fermi energy of ferromagnets, $E_{\rm F}^{(f)}=(E_{\rm
F}^{\uparrow }+E_{\rm F}^{\downarrow })/2$. Moduli of the Fermi
wave vectors, $k_{\rm F}^{(s)}=\sqrt{2mE_{\rm F}^{(s)}/\hbar
^{2}}$ and $k_{\rm F}^{(f)}=\sqrt{2mE_{\rm F}^{(f)}/\hbar ^{2}}$
in S and F may be different in general.  The in-plane, $y$-$z$,
magnetizations of the neighboring F layers are not collinear in
general, and the magnetic domain structure is described by the
angle ${\alpha (}\mathbf{r)}$ with respect to the $z$-axes:
${\alpha (}\mathbf{r)}=-\alpha/2$ for $-d_{1}<x<0$ in F$_{1}$, and
${\alpha (}\mathbf{r)}=\alpha/2 $ for $0<x<d_{2}$ in F$_{2}$. We
assume equal magnitudes of the exchange interaction in
ferromagnetic domains, $h{(}\mathbf{r)}=h_{0}\Theta
(x+d_{1})\Theta (d_{2}-x)$, where $\Theta (x)$ stands for the
Heaviside step function. We also assume that the two
superconductors are identical, and neglecting self-consistency, we
take the pair potential $ \Delta (\mathbf{r})$ in the form
\begin{equation}
\Delta (\mathbf{r})=\Delta \left[ e^{-i\phi /2}\Theta
(-x-d_{1})+e^{i\phi /2}\Theta (x-d_{2})\right] ,
\end{equation}%
where $\Delta $ is the bulk superconducting gap and $\phi $ is the
macroscopic phase difference across the junction. The temperature
dependence of $\Delta $ is given by $\Delta (T)=\Delta (0)\tanh
\left( 1.74\sqrt{T_{c}/T-1}\right)$.\cite{munhl}

Note that self-consistency may be safely neglected when the
proximity effect is weak between  S and F layers. This includes
the situations with large tunnel barriers at interfaces, and/or
narrow F constriction, and/or large Fermi velocities mismatch. For
a planar junction geometry and good contacts between S and F
layers, $\Delta$ will be suppressed at the vicinity of FS
interfaces.

The parallel component of the wave vector, $\mathbf{k}_{||}$, is
conserved due to translational invariance of the junction in
directions perpendicular to the $x$-axis. Consequently, the wave
function can be written in the form
\begin{equation}
\Psi (\mathbf{r})=\psi(x)e^{i\mathbf{k}_{||}\cdot \mathbf{r}},
\end{equation}
where $\psi (x)=[u_{\uparrow }(x),u_{\downarrow }(x),v_{\downarrow
}(x),v_{\uparrow }(x)]^{\rm T}$ satisfies the boundary conditions
\begin{eqnarray}
\label{bc4}
\psi (x)|_{x_{i}+0} =\psi (x)|_{x_{i}-0}&=&\psi (x_{i}), \\
\frac{d\psi (x)}{dx}\Big|_{x_{i}+0}-\frac{d\psi
(x)}{dx}\Big|_{x_{i}-0} &=&Z_{i}k_{\rm F}^{(s)}\psi (x_{i}).
\end{eqnarray}
Here, $Z_{i}=2mW_{i}/\hbar ^{2}k_{\rm F}^{(s)}$ ($i=1,2,3$) are
parameters that measure the strength of each insulating interface
located at $x_{i}=-d_1$, $0$, $d_2$.

The four independent solutions of the scattering problem for
Eq.~(\ref {bgd0}) correspond to the four types of quasiparticle
injection processes: an electron-like
 or a hole-like quasiparticle  injected from either the left
or from the right superconducting electrode. \cite{Furusaki
Tsukada} When an electron-like quasiparticle is injected from the
left superconductor, the solutions of Eq.~(\ref {bgd0}) are
\begin{eqnarray}
\label{eqS1}
u_{\uparrow}(x)&=&(e^{ik^{+}x}+b_{\uparrow}e^{-ik^{+}x})\bar{u}e^{-i\phi
/2}\nonumber\\ ~&~&+a_{\uparrow }e^{ik^{-}x}\bar{v}e^{-i\phi /2},\\
u_{\downarrow}(x)&=&(e^{ik^{+}x}+b_{\downarrow}e^{-ik^{+}x})\bar{u}e^{-i\phi
/2}\nonumber\\ ~&~&+a_{\downarrow}e^{ik^{-}x}\bar{v}e^{-i\phi /2},\\
v_{\downarrow}(x)&=&(e^{ik^{+}x}+b_{\uparrow}e^{-ik^{+}x})\bar{v}
+a_{\uparrow }e^{ik^{-}x}\bar{u},\\
v_{\uparrow}(x)&=&(e^{ik^{+}x}+b_{\downarrow}e^{-ik^{+}x})\bar{v}
+a_{\downarrow}e^{ik^{-}x}\bar{u},
\end{eqnarray}%
for the left superconductor ($x<0$),
\begin{eqnarray}
\label{eqS2}
u_{\uparrow}(x)&=&c_{\uparrow}e^{ik^{+}x}\bar{u}e^{i\phi /2}+
d_{\uparrow}e^{-ik^{-}x}\bar{v}e^{i\phi /2},\\
u_{\downarrow}(x)&=&c_{\downarrow}e^{ik^{+}x}\bar{u}e^{i\phi /2}+
d_{\downarrow}e^{-ik^{-}x}\bar{v}e^{i\phi /2},\\
v_{\downarrow}(x)&=&c_{\uparrow}e^{ik^{+}x}\bar{v}+
d_{\downarrow}e^{-ik^{-}x}\bar{u},\\
v_{\uparrow}(x)&=&c_{\downarrow}e^{ik^{+}x}\bar{v}+
d_{\downarrow}e^{-ik^{-}x}\bar{u},
\end{eqnarray}%
for the right superconductor ($x>d_2$), and
\begin{eqnarray}
u_{\uparrow}(x)&=&C_1e^{iq_{\uparrow}^{+}x}\cos{(\alpha/2)}+
C_2e^{-iq_{\uparrow}^{+}x}\cos{(\alpha/2)}\nonumber\\ ~&~&-i
C_3e^{iq_{\downarrow}^{+}x}\sin{(\alpha/2)}-i
C_4e^{-iq_{\downarrow}^{+}x}\sin{(\alpha/2)},\nonumber\\ \\
u_{\downarrow}(x)&=&-iC_1e^{iq_{\uparrow}^{+}x}\sin{(\alpha/2)}-i
C_2e^{-iq_{\uparrow}^{+}x}\sin{(\alpha/2)}\nonumber\\ ~&~&+
C_3e^{iq_{\downarrow}^{+}x}\cos{(\alpha/2)}+
C_4e^{-iq_{\downarrow}^{+}x}\cos{(\alpha/2)},\nonumber\\ \\
v_{\downarrow}(x)&=&iC_5e^{iq_{\uparrow}^{-}x}\sin{(\alpha/2)}+i
C_6e^{-iq_{\uparrow}^{-}x}\sin{(\alpha/2)}\nonumber\\ ~&~&+
C_7e^{iq_{\downarrow}^{-}x}\cos{(\alpha/2)}+
C_8e^{-iq_{\downarrow}^{-}x}\cos{(\alpha/2)},\nonumber\\ \\
v_{\uparrow}(x)&=&C_5e^{iq_{\uparrow}^{-}x}\cos{(\alpha/2)}+
C_6e^{-iq_{\uparrow}^{-}x}\cos{(\alpha/2)}\nonumber\\ ~&~&+i
C_7e^{iq_{\downarrow}^{-}x}\sin{(\alpha/2)}+i
C_8e^{-iq_{\downarrow}^{-}x}\sin{(\alpha/2)}, \nonumber\\
\label{eqF}
\end{eqnarray}%
for the left ferromagnetic layer F$_1$ ($-d_1<x<0$). Solutions for
the right ferromagnetic layer F$_2$ ($0<x<d_2$) can be obtain by
substitution $\alpha\rightarrow -\alpha$, with a new set of
constants $C'_1,\dots,C'_8$. Here, $\bar{u}=\sqrt{(1+\Omega
/E)/2}$ and $\bar{v}=\sqrt{(1-\Omega /E)/2}$ are the BCS
amplitudes, and $\Omega =\sqrt{E^{2}-\Delta ^{2}}$. Longitudinal
($x$-) components of the wave vectors are
\begin{equation}
k^{\pm }=\sqrt{ (2m/\hbar ^{2})(E_{\rm F}^{(s)}\pm \Omega
)-\mathbf{k}_{||}^{2}}  \label{ksupra}
\end{equation}
in superconductors, and
\begin{equation}
q_{\sigma }^{\pm}=\sqrt{(2m/\hbar ^{2})(E_{\rm
F}^{(f)}+\rho_{\sigma} h_{0}\pm E)-\mathbf{k}_{||}^{2}}
\label{qferro}
\end{equation}
in ferromagnetic layers. The sign $+$ or $-$ in the superscript
corresponds to the sign of quasiparticle energy, whereas
$\rho_{\sigma}=+1(-1)$ is related to the spin projection
$\sigma=\uparrow$($\downarrow$).

Note that solutions for quasiparticles with opposite spin
orientation in F layers are coupled due to misorientation of
magnetizations. Solutions are decoupled for $\alpha=0$ and
$\alpha=\pi$. In these cases spin triplet superconducting
correlations with nonzero spin projections also vanish. The case
of transparent SFFS junctions with $\alpha=\pi$ was treated
previously by Blanter and Hekking.\cite{blanter4} For $\alpha=0$,
$Z_2=0$, and $Z_1=Z_3$ solutions reduce to analytic expressions
previously obtained for SIFIS junctions\cite{radovic03,cayssol04}
with a net thickness of the ferromagnetic layer $d=d_1+d_2$.

Complete solution of the scattering problem requires determination
of $24$ unknown coefficients: $4+4$ constants in superconducting
electrodes ($a_{\sigma }$ and $b_{\sigma }$ in the left S, and
$c_{\sigma }$ and $d_{\sigma }$ in the right S for two spin
orientations) and $8+8$ constants in ferromagnetic electrodes
($C_1,\dots,C_8$ in F$_1$ and $C_1^{'},\dots,C_8^{'}$ in F$_2$).
When applied to the solutions given by
Eqs.~(\ref{eqS1})--(\ref{eqF}), the boundary conditions at three
interfaces, Eq.~(\ref{bc4}), provide the necessary $24$ equations.
Analogously, one can find the solutions for other three channels
of quasiparticle injection processes. However, the Andreev
amplitudes in the first channel are sufficient to calculate the
Josephson current.\cite{Furusaki Tsukada}


\section{Josephson current}

The stationary Josephson current can be expressed in terms of the
Andreev reflection amplitudes, $a_{\sigma}=a_{\sigma}({\phi})$, by
using the temperature Green's function formalism\cite{Furusaki
Tsukada}
\begin{eqnarray}
I(\phi)&=&\frac{e\Delta }{2\hbar }\sum_{\sigma
,\mathbf{k}_{||}}k_{B}T \times\nonumber\\
&~&\times\sum_{\omega _{n}}\frac{1}{2\Omega
_{n}}(k_{n}^{+}+k_{n}^{-})\left( \frac{a_{\sigma n}(\phi
)}{k_{n}^{+}}-\frac{a_{\sigma n}(-\phi )}{k_{n}^{-}}\right), \nonumber\\
\end{eqnarray}%
where $k_{n}^{+},~k_{n}^{-}$, and $a_{\sigma n}(\phi )$ are
obtained from $k^{+},~k^{-}$, and $a_{\sigma }(\phi )$ by the
analytic continuation $E\rightarrow i\omega _{n}$. The Matsubara
frequencies are $\omega _{n}=\pi k_{B}T(2n+1)$, $n=0,\pm 1,\pm
2,...$, and $\Omega _{n}=\sqrt{\omega _{n}^{2}+\Delta ^{2}}$.

For simplicity, we illustrate our results for a single transverse
channel, ${\bf k}_{\|}=0$, symmetric junctions, $d_{1}=d_{2}=d/2$,
and equal Fermi wave vectors, $k_{\rm F}^{(s)}=k_{\rm
F}^{(f)}=k_{\rm F}$. Superconductors are characterized by fixed
bulk value of the pair potential, $\Delta /E_{\rm
F}^{(s)}=10^{-3}$. In order to adhere to realistic parameters  all
the examples are shown for $k_{\rm F}d<100$ since it still seems
unlikely that the ballistic limit can be experimentally achieved
for thicker ferromagnets.

\begin{figure}[h]
\begin{center}
    \includegraphics[width=7cm]{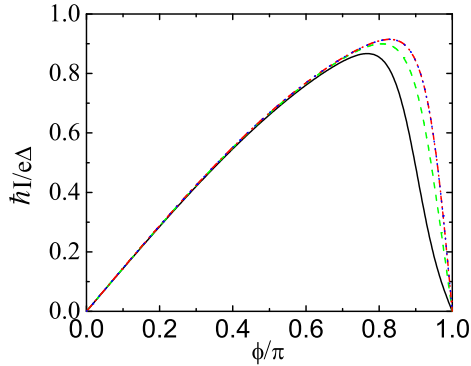}
\caption{(Color online) The current-phase relation $I(\phi)$ for
$T/T_c=0.1$, $h_0/E_{\rm F}=0.01$, $Z_1=Z_2=Z_3=0$, $dk_{\rm
F}=30$, and three different values of the misorientation angle:
$\alpha=0$ (solid curve), $\alpha=\pi/2$ (dashed curve), and
$\alpha=\pi$ (dash-dotted curve). The latter coincides with
$I(\phi)$ for the corresponding SNS junction ($h_0=0$).}
\label{Fig2}
\end{center}
\end{figure}
\begin{figure}[h]
\begin{center}
    \includegraphics[width=7cm]{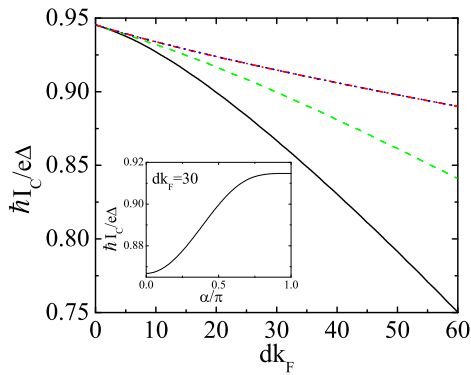}
\caption{(Color online) The critical current $I_c$ as a function
of $d$ for $T/T_c=0.1$, $h_0/E_{\rm F}=0.01$, $Z_1=Z_2=Z_3=0$, and
three different values of the misorientation angle: $\alpha=0$
(solid curve), $\alpha=\pi/2$ (dashed curve), and $\alpha=\pi$
(dash-dotted curve). The latter approximately coincides with
$I_c(d)$ for the corresponding SNS junction (dotted curve). Inset:
$I_c$ as a function of $\alpha$ for $dk_{\rm F}=30$.} \label{Fig3}
\end{center}
\end{figure}
\begin{figure}[h]
\begin{center}
    \includegraphics[width=7cm]{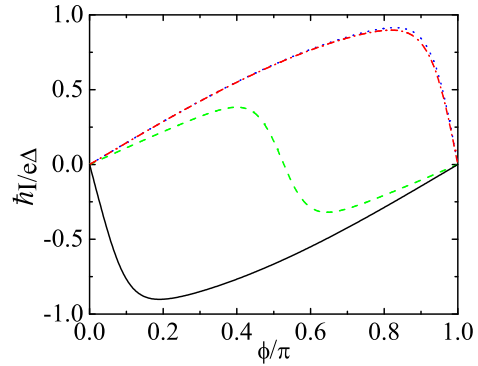}
\caption{(Color online) The current-phase relation $I(\phi)$ for
$T/T_c=0.1$, $h_0/E_{\rm F}=0.1$, $Z_1=Z_2=Z_3=0$, $dk_{\rm
F}=30$, and three different values of the misorientation angle:
$\alpha=0$ (solid curve), $\alpha=\pi/2$ (dashed curve), and
$\alpha=\pi$ (dash-dotted curve). The latter approximately
coincides with $I_c(d)$ for the corresponding SNS junction (dotted
curve).} \label{Fig4}
\end{center}
\end{figure}
\begin{figure}[h]
\begin{center}
    \includegraphics[width=7cm]{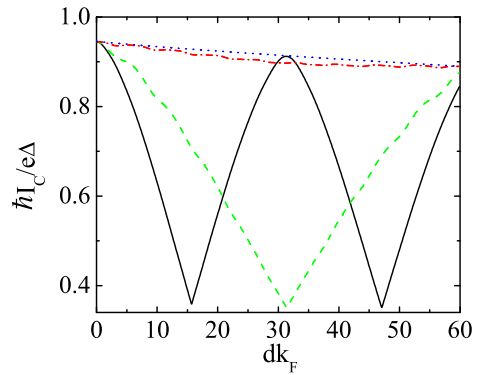}
\caption{(Color online) The critical current $I_c$ as a function
of $d$ for $T/T_c=0.1$, $h_0/E_{\rm F}=0.1$, $Z_1=Z_2=Z_3=0$, and
three different values of the misorientation angle: $\alpha=0$
(solid curve), $\alpha=\pi/2$ (dashed curve), and $\alpha=\pi$
(dash-dotted curve). Dotted curve represents $I_c(d)$ for the
corresponding SNS junction ($h_0=0$).}\label{Fig5}
\end{center}
\end{figure}
\begin{figure}[h]
\begin{center}
    \includegraphics[width=7cm]{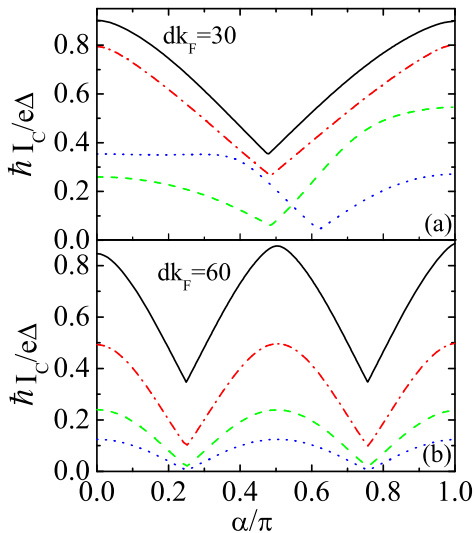}
\caption{(Color online) The critical current $I_c$ as a function
of the misorientation angle $\alpha$ for $T/T_c=0.1$, $h_0/E_{\rm
F}=0.1$, and four sets of interface transparencies:
$Z_1=Z_2=Z_3=0$ (solid curve), $Z_1=Z_3=0$, $Z_2=1$ (dash-dotted
curve), $Z_1=Z_3=1$, $Z_2=0$ (dashed curve), and $Z_1=Z_2=Z_3=1$
(dotted curve). Panel (a): $dk_{\rm F}=30$. Panel (b): $dk_{\rm
F}=60$. The dips in $I_c(\alpha)$ separate alternating $0$ and
$\pi$ states, beginning with a $0$ state from the left.}
\label{Fig6}
\end{center}
\end{figure}
\begin{figure}[h]
\begin{center}
    \includegraphics[width=7cm]{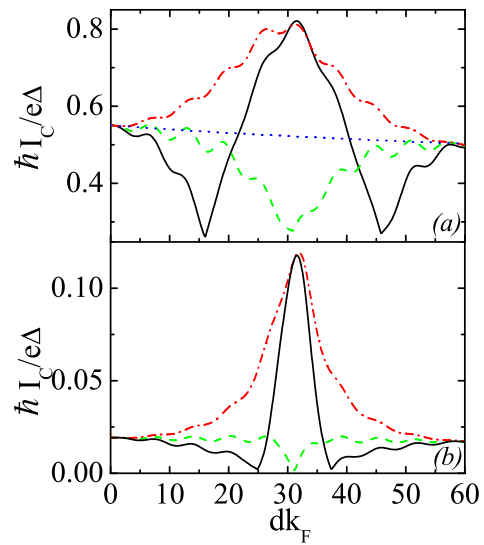}
\caption{(Color online) The critical current $I_c$ as a function
of $d$ for an SFIFS junction with $h_0/E_{\rm F}=0.1$ at
$T/T_c=0.1$, and three different values of the misorientation
angle: $\alpha=0$ (solid curve), $\alpha=\pi/2$ (dashed curve),
and $\alpha=\pi$ (dash-dotted curve). Panel (a): $Z_1=Z_3=0$,
$Z_2=1$. Panel (b): $Z_1=Z_3=0$, $Z_2=10$. $I_c(d)$ for the
corresponding SNINS junction (dotted curves) are shown for
comparison.} \label{Fig7}
\end{center}
\end{figure}

For fully transparent interfaces, $Z_{1}=Z_{2}=Z_{3}=0$, and for a
weak exchange field, $h_{0}/E_{\rm F}=0.01$, the current-phase
relations are shown in Fig.~\ref{Fig2} for three different values
of misorientation angle $\alpha =0$, $\pi /2$, $\pi$. The figure
illustrates  a junction with thin ferromagnetic interlayer,
$k_{\rm F}d=30$, at low temperature $ T/T_{c}=0.1$. As expected,
the $I(\phi)$ curve for $\alpha =\pi$ is the same as for the
corresponding SNS junction ($h_0=0$) because the influence of
opposite magnetizations in $F_{1}$ and $F_{2}$ practically cancels
out. The $I(\phi )$ relations are anharmonic due to full
transparency and low temperature. The junction is clearly in the
$0$-state for all values of misorientation angle; for such a small
$h_0$ there are no oscillations of the critical current related to
the $0-\pi$ transitions for $k_{\rm F}d<100$, Fig.~\ref{Fig3}. We
have shown that the critical current increases monotonicaly with
increasing $\alpha$, from the SFS  ($\alpha =0$) to the SNS
($\alpha =\pi$) value. It is evident that spin triplet
correlations, present for $0<\alpha<\pi$, here do not induce a
non-monotonic variation of $I_c(\alpha)$.  For larger exchange
energy, $h_0/E_{\rm F}=0.1$, the current-phase relations are shown
in Fig.~\ref{Fig4} for the same values of $\alpha$, $k_{\rm F}d$,
and $T/T_c$ as in Fig.~\ref{Fig2}. In Fig.~\ref{Fig5}, for $\alpha
=\pi/2$ the $I(\phi)$ curve exhibits coexistence of $0$ and $\pi$
states with a dominant contribution of the second
harmonic.\cite{radovic01} The critical Josephson current
oscillates as a function of the ferromagnetic interlayer thickness
due to the onset of $0-\pi$ transitions, except for $\alpha$ close
to $\pi$. The period of oscillations increases with $\alpha$. This
is simply the effect of a weaker influence of the exchange
potential in the case of non-parallel magnetizations. For example,
the period of oscillations in $I_c$ is approximately twice bigger
for $\alpha =\pi/2$ than for $\alpha=0$. We have obtained the same
result for parallel, but twice smaller, exchange energy
($\alpha=0$ and $h_0/E_{\rm F}=0.05$). It can be shown (from the
calculation of the junction free energy) that the dips in $I_c(d)$
curves correspond to $0-\pi$ transitions. Similar oscillations of
the critical Josephson current as a function of $\alpha$ for fixed
$k_{\rm F}d$ (either $30$ or $60$) are shown in Fig.~\ref{Fig6},
as well as the influence of finite interfacial transparency. As
already suggested for diffusive
junctions,\cite{bergeret03prbtriplet,golubov2} it can be clearly
seen that the transition between $0$ and $\pi$ states can be
induced by varying the misorientation angle $\alpha$. This feature
is of particular importance for experimental applications, since
fine tuning of $0-\pi$ transitions could be realized more easily
by varying $\alpha$ than by varying thickness or temperature.

The influence of geometrical resonances is illustrated in
Fig.~\ref{Fig7} for $Z_{1}=Z_{3}=0$, and (a) $Z_{2}=1$ and (b)
$Z_{2}=10$. All other parameters are the same as in
Fig.~\ref{Fig5}. Oscillations of the critical current due to
geometrical resonances are superimposed on the oscillations
related to the transitions between $0$ and $ \pi $ states. This
effect is clearly visible in Fig.~\ref{Fig7}(b) for low
transparency of the interface between magnetic domains
($Z_{2}=10$). We emphasize that now, due to the presence of a
barrier between the F layers, the magnetic influence persists for
$\alpha=\pi$, and $I_c(d)$ considerably differs from the results
obtained for the corresponding SNINS junction. However, for
identical domains, there are still no transitions to the $\pi$
state in the AP configuration. It can be seen that the amplitudes
of geometrical oscillations of the supercurrent are significantly
larger in SFIFS than in the corresponding SNINS junctions (dotted
curves in Fig.~\ref{Fig7}). Rapid oscillation of $I_c(d)$  can be
also seen in the case of finite transparency of interfaces between
ferromagnetic layers and superconductors,\cite{radovic03} for any
value of $\alpha$. This is due to the resonant amplification of
the Josephson current by quasi-bound states entering the
superconducting gap as the thickness $d$ of the interlayer is
varied.\cite{ivana}  The effect of finite transparency of all the
interfaces is similar. In a more realistic case of planar
(multichannel) junctions these geometrical oscillations of the
critical current are damped, while positions of maxima and minima
are slightly shifted.\cite{radovic03}

\begin{figure}[h]
\begin{center}
    \includegraphics[width=7cm]{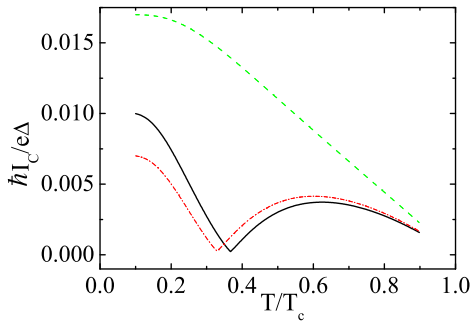}
\caption{(Color online) The critical current $I_c$ as a function
of $T/T_c$ for $dk_{\rm F}=43$, $h_0/E_{\rm F}=0.1$,  $Z_1=Z_3=0$,
$Z_2=10$ and three different values of the misorientation angle:
$\alpha=0$ (solid curve), $\alpha=\pi/16$ (dash-dotted curve), and
$\alpha=\pi/2$ (dashed curve). For $\alpha\approx 0$ the
characteristic nonmonotonic variation is related to the transition
from $0$ state (low $T$) to $\pi$ state (higher $T$) with dip at
the transition. For $\alpha=\pi/2$ the junction is in $0$ state at
all temperatures.}\label{Fig8}
\end{center}
\end{figure}

Transitions between $0$ and $\pi$ states can be induced by
changing the temperature of a junction with low transparency and
strong ferromagnetic influence.\cite{radovic03,nikolaj,barash02}
This is illustrated in Fig.~\ref{Fig8} for $Z_{1}=Z_{3}=0$, and
$Z_{2}=10$, $k_{\rm F}d=43$, $h_{0}/E_{\rm F}=0.1$ and for three
values of $ \alpha =0$, $\pi /16$, $\pi/2$. It can be seen that
temperature dependence is very sensitive to the value of $\alpha$;
for example, the temperature induced transition from $0$ to $\pi$
state when $\alpha\approx 0$ is absent for $\alpha=\pi/2$.

\section{Conclusion}
We have studied the Josephson effect in clean
superconductor-ferromagnet heterojunctions containing two
mono-domain ferromagnetic layers with arbitrary transparency of
the interfaces and any angle $\alpha$ between magnetizations,
including two previously considered limiting cases $\alpha=0$ and
$\pi$.~\cite{radovic03,blanter4}  The Josephson current is
calculated numerically via Bogoliubov--de Gennes formalism. We
have found that transitions between $0$ and $\pi$ states,
resulting in characteristic dips in $I_c(\alpha)$ curves, can be
induced by varying the relative orientation of magnetizations,
like in diffusive junctions.~\cite{bergeret03prbtriplet,golubov2}
However, this is simply the effect of a weaker influence of the
exchange potential in the case of non-parallel magnetizations. No
substantial impact of spin-triplet superconducting correlations on
the Josephson current has been found in the clean limit. For weak
ferromagnets, far from $0-\pi$ transitions, the critical Josephson
current monotonically depends on the angle between magnetizations.
While in fully transparent junctions oscillatory dependence of the
critical Josephson current on junction parameters is related only
to the $0-\pi$ transitions, for finite transparency of interfaces
pronounced geometrical oscillations occur due to coherent
contribution of quasiparticle transmission resonances (quasi-bound
states) to the Andreev process.

In conclusion, due to recent progress in nanofabrication
techniques,~\cite{ivan} the SI${_1}$F${_1}$F${_2}$I${_2}$S
junction may be realized in the clean regime in a setup where the
angle $\alpha$ could be tuned by a weak external magnetic field.
We thus suggest experimental investigation of the predicted
control of the critical current and $0-\pi$ transitions by varying
the angle between magnetizations in Josephson junctions with
magnetic bilayers.

\bigskip


\section{Acknowledgment}

The work was supported in part by French ECO-NET EGIDE program,
and by the Serbian Ministry of Science, Project No.~141014.

\end{document}